\documentclass{article}  

\usepackage{amsmath}
\usepackage{amsfonts}
\usepackage{amssymb}
\usepackage{amsthm}
\usepackage{mathtools}
\usepackage{mathrsfs}
\usepackage{mathpazo}

\usepackage{graphicx}
\usepackage{enumerate,color}
\usepackage{microtype}
\usepackage[margin=15pt,font=small,labelfont={bf,sf}]{caption}
\usepackage{sectsty}
\allsectionsfont{\sffamily}
\usepackage[parfill]{parskip}

\usepackage{tikz}
\usepackage{pgfplots}
\usepgfplotslibrary{groupplots}
\pgfplotsset{compat=newest}
\tikzset{every picture/.style=thick}
\usetikzlibrary{shapes,backgrounds,calc,arrows}

\usepackage{hyperref}
\usepackage[nameinlink]{cleveref}
\crefformat{equation}{(#2#1#3)}
\crefmultiformat{equation}{(#2#1#3)}%
{ and~(#2#1#3)}{, (#2#1#3)}{ and~(#2#1#3)}

\definecolor{lightblue}{rgb}{0.54, 0.81, 0.94}
\hypersetup{
    colorlinks = true,
    linkcolor = lightblue,
    citecolor = lightblue,
    urlcolor = lightblue,
    }

\newcommand{\cfunof}[1]{\ensuremath{\left\{#1\right\}}}
\newcommand{\funof}[1]{\ensuremath{\left(#1\right)}}

\DeclareMathOperator{\proj}{proj_{\mathcal{C}}}

\DeclareMathOperator{\tr}{tr}

\newtheorem{theorem}{Theorem}
\newtheorem{remark}{Remark}
\crefname{remark}{remark}{Remark}

\begin{document}

\title{Lyapunov equations: a (fixed) point of view}
\author{Richard Pates\thanks{The author is a member of the ELLIIT Strategic Research Area at Lund University. This work was supported by the ELLIIT Strategic Research Area. This project has received funding from ERC grant agreement No 834142.}}

\maketitle

\begin{abstract}
The Lyapunov equation is the gateway drug of nonlinear control theory. In these notes we revisit an elegant statement connecting the concepts of asymptotic stability and observability, to the solvability of Lyapunov equations, and discuss how this statement can be proved using the Brouwer fixed-point theorem.
\end{abstract}

%%%%%%%%%%%%%%%%%%%%%%%%%%%%%%%%%%%%%%%%%%%%%%%%%%%%%%%%%%%%%%%%%%%%%%%%%%%%%%%%

\section{The statement}

\begin{figure}
\centering
\begin{tikzpicture}
  \node[draw, regular polygon, regular polygon sides=3, minimum size=5cm, label={[label distance=-1.5cm]below:(i)}] (i) at (0.5,0) {};
  \node[draw, regular polygon, regular polygon sides=3, minimum size=5cm, label={[label distance=-1.5cm]below:(ii)}] (ii) at (3.5,0) {};
  \node[draw, regular polygon, regular polygon sides=3, minimum size=5cm, label={[label distance=-1.5cm]below:(iii)}] (iii) at (2,-{sqrt(27)/2}) {};
  % filling intersections
  \begin{scope}[on background layer]
    \fill[blue!30, opacity=0.3] (i.corner 1) -- (i.corner 2) -- (i.corner 3) -- cycle;
    \fill[red!30, opacity=0.3] (ii.corner 1) -- (ii.corner 2) -- (ii.corner 3) -- cycle;
    \fill[green!30, opacity=0.3] (iii.corner 1) -- (iii.corner 2) -- (iii.corner 3) -- cycle;
    \clip (i.corner 1) -- (i.corner 2) -- (i.corner 3) -- cycle;
    \clip (ii.corner 1) -- (ii.corner 2) -- (ii.corner 3) -- cycle;
    \fill[yellow!30, opacity=0.3] (iii.corner 1) -- (iii.corner 2) -- (iii.corner 3) -- cycle;
  \end{scope}
  % Draw a bounding box
  \draw ([shift={(-0.5,-0.5)}]current bounding box.south west) rectangle ([shift={(0.5,0.5)}]current bounding box.north east);
\end{tikzpicture}
\caption{Euler diagram showing the interdependence of the statements in \Cref{thm:main}. The theorem shows that it is not possible for any pair of the statements to be true without the third also being true.}\label{fig:1}
\end{figure}
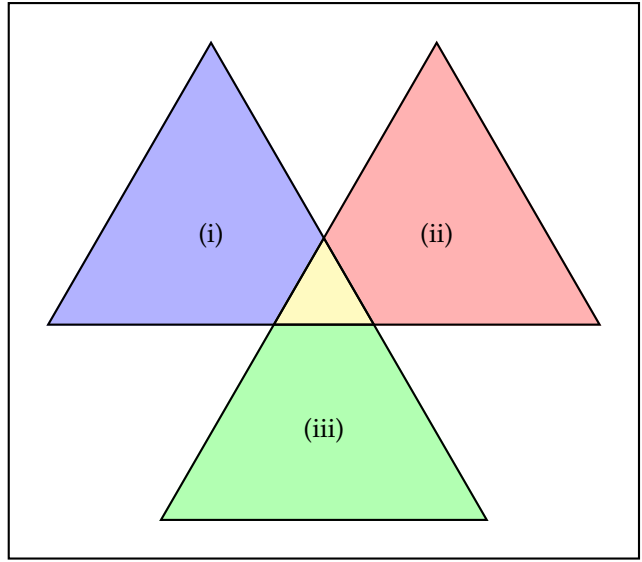

I would like to present and (partially) prove an interesting statement about the existence of solutions to the discrete time Lyapunov equation
\begin{equation}\label{eq:lyap}
A^{\mathsf{T}}QA-Q+C^{\mathsf{T}}C=0.
\end{equation}
As usual, the matrices $A$ and $C$ come from the discrete time dynamical system
\begin{equation}\label{eq:dynsys}
x_{k+1}=Ax_k,\;y_k=Cx_k
\end{equation}
and we are interested in positive definite matrices $Q$ such that \cref{eq:lyap} holds. 

Our focus will be the following theorem, connecting the concepts of stability, observability, and the solvability of \cref{eq:lyap} with a positive definite $Q$. An informal survey of the faculty at Lund University revealed that the statement we are about to see is not as well known as it deserves to be. It was first brought fully into my consciousness by Laurent Lessard (Northeastern University) when discussing undergraduate teaching during CDC 2022. I remember being very enthusiastic at the time. However, upon my return, I rather embarrassingly discovered that this was precisely the statement given in an undergraudate control course that I had both taken (and even worse, taught) during my time at Cambridge. Naturally analogous versions can be given in continuous time, and also involving controllability rather than observability (these, as well as variants based on detectability/stabilisability are briefly discussed in \Cref{rem:1}). But without further ado, here is the theorem.

\begin{theorem}\label{thm:main}
Given a discrete time dynamical system
\[
\begin{aligned}
x_{k+1}&=Ax_k,\,y_k=Cx_k
\end{aligned}
\]
any two of the following conditions imply the third:
\begin{enumerate}
\item[(i)] The discrete time dynamical system is asymptotically stable.
\item[(ii)] The discrete time dynamical system is observable.
\item[(iii)] There exists a (unique) positive definite solution to the Lyapunov equation
\[
A^{\mathsf{T}}QA-Q+C^{\mathsf{T}}C=0.
\]
\end{enumerate}
\end{theorem}
This theorem combines many of the usual statements about Lyapunov equations into a single compact package. For example, the statement ”an observable dynamical system as in \cref{eq:dynsys} is asymptotically stable if and only if there exists a positive definite solution to the Lyapunov equation in \cref{eq:lyap},” follows from the implications
\[
\funof{\text{i}}\land{}\funof{\text{ii}}\implies{}\funof{\text{iii}}\;\;\text{and}\;\;\funof{\text{ii}}\land{}\funof{\text{iii}}\implies{}\funof{\text{i}}.
\]
Combining implications in a similar fashion shows that for an asymptotically stable system, observability may be certified from the solution to a Lyapunov equation (in this context, the resulting $Q$ is normally called the observability Gramian), and if we assume (iii), then asymptotic stability and observability are equivalent!

\begin{remark}\label{rem:1}
The version of \Cref{thm:main} involving controllability states that given a discrete-time dynamical system
\[
\begin{aligned}
x_{k+1}&=Ax_k+Bu_k
\end{aligned}
\]
any two of the following conditions imply the third:
\begin{enumerate}[(i)]
\item The discrete time dynamical system is asymptotically stable.
\item The discrete time dynamical system is controllable.
\item There exists a (unique) positive definite solution to the Lyapunov equation
\[
A^{\mathsf{T}}PA-P+BB^{\mathsf{T}}=0.
\]
\end{enumerate}
Many other stability theorems involving Lyapunov equations are known (see, for example, \cite[Chapter 7]{PW98}). These often revolve around relaxations of (i)--(iii). For example, it is known that if conditions (ii) and (iii) in \Cref{thm:main} are weakened to:
\begin{enumerate}
\item[(ii)'] The discrete time dynamical system is \textit{detectible};
\item[(iii)'] There exists a (unique) positive-\textit{semi}definite solution to the Lyapunov equation
\[
A^{\mathsf{T}}QA-Q+C^{\mathsf{T}}C=0;
\]
\end{enumerate}
then together, $\text{(ii)'}$ and $\text{(iii)'}$ imply (i) (asymptotic stability). That is, when using Lyapunov equations to test for stability, stronger results than \Cref{thm:main} are available. However, theorems based on these other statements tend to lose the appealing symmetry of \Cref{thm:main}, which is why we have highlighted this result in these notes (with respect to (ii)' and (iii)', $\text{(i)}\implies{}\text{(ii)'}$ and $\text{(i)}\implies{}\text{(iii)'}$, so in fact,  $\text{(i)}\iff{}\text{(ii)'}\land\text{(iii)'}$). The proof given in these notes can be adapted to cover the implication $\text{(i)}\implies{}\text{(iii)'}$ (although the argument given in \cref{sec:unique} must be strengthened). Can you figure out the details?
\end{remark}
 
\section{The proof}

We will now turn our attention to proving \Cref{thm:main}, or rather, proving the implication
\[
\text{(i)}\land\text{(ii)}\implies{}\text{(iii)}
\]
since this is (in my opinion) the most interesting. The remaining implications are briefly discussed in \Cref{rem:2}. We will take a slightly unconventional approach that relies on very little linear algebra. This both emphasizes the ``elementary'' nature of the theorem (in the sense that it follows from more basic mathematical facts, rather than that the proof will be easy...) and provides a little variety, compared to standard proofs centered on gadgets like Jordan normal forms and Kronecker products. It also serves to illustrate how to generalize the theorem to other settings (and implicitly that the nice linear algebra algorithms also often generalize) and sets the stage for much of the nonlinear control theory that has taken inspiration from this result (see \Cref{rem:3} for an example of a positive systems analog of \Cref{thm:main} that follows immediately from the presented arguments). 

\subsection{Applying the Brouwer fixed-point theorem}

Our basic approach will be to view the Lyapunov equation as a fixed-point equation. This will allow us to use the Brouwer fixed-point theorem to deduce existence of solutions of the required form. To the uninitiated, the version of the Brouwer fixed-point theorem that we will require reads as follows.

\begin{theorem}[The Brouwer fixed-point theorem]\label{thm:brouwer}
Let $\mathcal{C}$ be a compact convex set. If $f:\mathcal{C}\rightarrow{}\mathcal{C}$ is a continuous function, then there exists a point $x\in\mathcal{C}$ such that $f\funof{x}=x$.
\end{theorem}

The Brouwer fixed-point theorem is famous both for its widespread use in many diverse areas of mathematics, as well as its many proofs. A surprising but very accessible proof comes from a combinatorial argument about coloring the vertices of triangles known as Sperner's lemma \cite{AZ18}, though equally surprising perhaps is that it is equivalent to the determinacy theorem for the game of Hex \cite{Gal79}! Its history is also rather convoluted, with roughly equivalent statements appearing in earlier works from Poincar\'{e}, Hadamard, and Bohl, as well as it becoming the subject of an intense feud between Cinquini and Scorza-Dragoni in the Italian mathematics community in the 1940s \cite{DM21}. 

We will make use of the Brouwer fixed-point theorem in a similar fashion to it's use in proofs of the Perron-Frobenius theorem \cite{Mac00}. I remember being quite struck by this line of reasoning when I first encountered it while browsing Pablo Parrilo's thesis \cite{Par00} (and procrastinating from writing my own), and I hope you will derive similar enjoyment if you haven't seen the argument before. Roughly speaking, we will recast the Lyapunov equation as the solution to an equation of the form
\[
x=\proj{} Mx+c
\]
where the right-hand side denotes the projection of the affine function $Mx+c$ onto a suitably chosen compact convex set $\mathcal{C}$. Under mild conditions (that are satisfied in our case),
\begin{equation}\label{eq:proj}
\proj{} Mx+c:\mathcal{C}\rightarrow{}\mathcal{C}
\end{equation}
is continuous on $\mathcal{C}$, meaning that the Brouwer fixed-point theorem ensures the existence of solutions of the Lyapunov equation. Using (i) and (ii) to take care of a few details to ensure things like uniqueness will then lead us to the conclusions of (iii). 

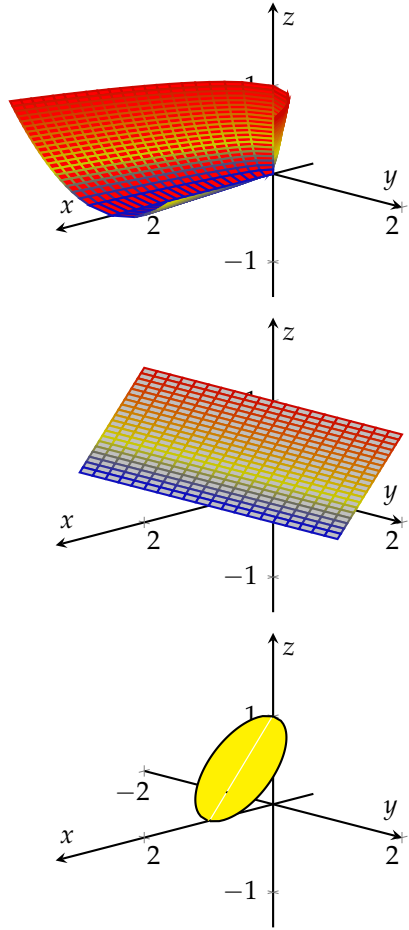
\begin{figure}
\centering
\begin{tikzpicture}
\begin{groupplot}[
    group style={
        group size=1 by 3,
        vertical sep=-1.5cm
    },
    	xlabel={$x$},
        ylabel={$y$},
        zlabel={$z$},
        zlabel style={rotate=-90},
        view={135}{15},
        xmin=-0.25,
        xmax=3,
        ymin=-2,
        ymax=2,
        zmin=-.25,
        zmax=.8,
        axis lines=center,
        axis equal,
	]
\nextgroupplot % the first plot
\addplot3[
    surf,
    domain=0:2.5,   % range for x
    domain y=0:1, % range for z
    samples=21,
    opacity=0.5,
    fill=red,
] 
({x}, {sqrt(abs(x*y))}, {y}); % equation of the surface y=sqrt(x*z)
\addplot3[
    surf,
    domain=0:2.5,   % range for x
    domain y=0:1, % range for z
    samples=21,
    opacity=0.5,
    fill=red,
] 
({x}, {-sqrt(abs(x*y))}, {y}); % equation of the surface y=-sqrt(x*z)
\nextgroupplot % the second plot
\addplot3[
    surf,
    domain=-0:1,   % range for x
    domain y=-2:2, % range for y
    samples=21,
    opacity=0.5,
    fill=lightgray,
] 
{1-x}; % equation of the surface x+z=1
\nextgroupplot % the third plot
\fill[
    domain=0:1,
    samples=50,
    color=yellow,
    opacity=0.5
] plot ({\x}, {sqrt(abs((\x-\x^2))}, {1-\x}) -- cycle;
\fill[
    domain=0:1,
    samples=50,
    color=yellow,
    opacity=0.5
] plot ({1-\x}, {-sqrt(abs(\x-\x^2))}, {\x}) -- cycle;
\draw[
    domain=0:1,
    samples=50,
    color=black,
    thick
] plot ({\x}, {sqrt(abs(\x-\x^2))}, {1-\x})
plot ({\x}, {-sqrt(abs(\x-\x^2))}, {1-\x});
\end{groupplot}
\end{tikzpicture}
\caption{Points on the red surface and above in the top axis satisfy the inequalities $xz\geq{}y^2$, $x\geq{}0$, and $z\geq{}0$. Points on the surface in the middle axis satisfy $x+z=1$. The circle on the third axis shows the intersection of these two shapes. The circle corresponds to the compact convex set $\mathcal{C}$ for the two by two matrices as defined by \cref{eq:unittr}. Points on this circle correspond to values of $x$, $y$, and $z$ such that the matrix in \cref{eq:2x2} is positive semi-definite with unit trace, and the function $f\funof{\cdot}$ in \cref{eq:fixptorig} maps points on this set back onto itself. Since the set is compact and convex, the Brouwer fixed-point theorem guarantees the existence of at least one fixed-point for this function (that is, a point on the circle that is mapped back onto itself by $f\funof{\cdot}$).}
\label{fig:2}
\end{figure}

We now proceed with the argument in detail (we return to the more abstract perspective in \Cref{rem:3}). Consider the set of positive semi-definite matrices with unit trace
\begin{equation}\label{eq:unittr}
\mathcal{C}=\cfunof{X:X\succeq{}0,\tr{X}=1}
\end{equation}
and the function $f:\mathcal{C}\rightarrow{}\mathcal{C}$ given by
\begin{equation}\label{eq:fixptorig}
f\funof{X}=\frac{A^{\mathsf{T}}XA+\alpha{}C^{\mathsf{T}}C}{\tr{\funof{A^{\mathsf{T}}XA+\alpha{}C^{\mathsf{T}}C}}},
\end{equation}
where $\alpha>0$ is a parameter. The set $\mathcal{C}$ for two by two matrices
\begin{equation}\label{eq:2x2}
X=\begin{pmatrix}x&y\\y&z\end{pmatrix}
\end{equation}
is illustrated in \Cref{fig:2}. Since $f\funof{\cdot}$ is continuous and $\mathcal{C}$ is compact and convex, \Cref{thm:brouwer} guarantees that for every $\alpha>0$, there exists a positive semi-definite $X_\alpha$ such that $X_\alpha=f\funof{X_\alpha}$. This gets us most of the way to finding a solution to Lyapunov equation. More specifically, denoting 
\[
\lambda_\alpha{}=\tr{\funof{A^{\mathsf{T}}X_\alpha{}A+\alpha{}C^{\mathsf{T}}C}},
\]
it then follows that
\[
X_\alpha=f\funof{X_\alpha}\;\implies\;\tfrac{\lambda_{\alpha}}{\alpha}X_{\alpha}=\funof{A^{\mathsf{T}}\tfrac{1}{\alpha}X_\alpha{}A+C^{\mathsf{T}}C}.
\]
Therefore, if we can show that for some $\alpha>0$,
\begin{enumerate}
\item[1)] $X_{\alpha}$ is positive definite,
\item[2)] $X_{\alpha}$ is unique, and
\item[3)] $\lambda_\alpha=1$;
\end{enumerate}
we will have shown the required implication by setting $Q=\tfrac{1}{\alpha}X_\alpha$. We will now explain how to use (i) and (ii) to deduce that there exists an $\alpha>0$ such that 1)--3) hold. 

\begin{remark}\label{rem:3}

The argument we are about to present works almost without change in the more abstract setting hinted at in \cref{eq:proj}. To get a flavour of how this works, observe that the numerator of $f\funof{\cdot{}}$ in \cref{eq:fixptorig} defines an affine function on the positive semi-definite cone. The effect of normalizing by the trace is then to project the output of this function, which will be a positive semi-definite matrix, onto a compact slice of the semi-definite cone (namely $\mathcal{C}$). In fact, the remaining steps of our proof continue to hold in the more abstract setting, allowing for analogs of Lyapunov equations and \Cref{thm:main} to be obtained for dynamical systems that evolve on other cones. 

To give a concrete example of such a generalization, suppose that we have the discrete time dynamical system
\[
x_{k+1}=Ax_k,\,y_k=cx_k
\]
where $A\geq{}0$ and $c\geq{}0$ (all the entries of $A$ and $c$ are non-negative), $c$ is a row vector (the output has dimension 1), and the initial condition is assumed to also be non-negative ($x_0\geq{}0$). Such systems are typically called internally positive systems in the literature \cite{RV18}. This is because they preserve positivity in the sense that the state trajectories $x_0,x_1,x_2,\ldots{}$ and output trajectories $y_0,y_1,y_2,\ldots{}$ remain non-negative. Since the non-negative orthant is itself a cone,  this system property motivates replacing $\mathcal{C}$ in \cref{eq:unittr} with a slice of non-negative orthant, namely the unit simplex $\mathcal{C}=\cfunof{x\geq{}0:\sum_{k=1}^nx_k=1}$, and $f\funof{\cdot{}}$ in  \cref{eq:fixptorig} with
\[
f\funof{x}=\frac{xA+\alpha{}c}{\sum_{k=1}^n\funof{xA}_k+\alpha{}c_k}.
\]
The ideas behind the presented proof work under these substitutions more or less without change [some of the algebra needs to be adjusted; for example, recalculating \cref{eq:xachain}, where in addition $\lambda_\alpha$ is set to ${\sum_{k=1}^n\funof{xA}_k+\alpha{}c_k}$]. Doing so leads to the conclusion that asymptotic stability and observability imply the existence of a $q>0$ (meaning that $q$ is entrywise positive) such that
\[
q=c+qA.
\]
This means that when the dynamical system is positive, condition (iii) can be replaced with the condition (iii)'' $c\funof{I-A}^{-1}>0$ [note the other implications can also be proved as hinted at in \Cref{rem:2}, but using the Lyapunov candidate $V\funof{x}=c\funof{I-A}^{-1}x$, so we really can replace (iii) with (iii)'' in \Cref{thm:main} in the internally positive system case].
\end{remark}

\subsection{Observability implies positive definiteness}

We will now deduce that 1) holds. Since $X_\alpha$ is a fixed-point, we can pass it through \cref{eq:fixptorig} as many times as we like and still obtain $X_\alpha$. This means that
\begin{align}
\nonumber{}X_\alpha=f^{\funof{n+1}}\funof{X_\alpha}&=f^{\funof{n}}\funof{\tfrac{1}{\lambda_\alpha}A^{\mathsf{T}}X_{\alpha}A+\tfrac{\alpha}{\lambda_\alpha}C^{\mathsf{T}}C}\\
\nonumber{}&=f^{\funof{n-1}}\funof{\tfrac{1}{\lambda_\alpha^2}\funof{A^2}^{\mathsf{T}}X_{\alpha}A^2+\tfrac{\alpha}{\lambda_\alpha^2}\funof{AC}^{\mathsf{T}}CA+\tfrac{\alpha}{\lambda_\alpha}C^{\mathsf{T}}C}\\
\nonumber{}&\,\,\,\vdots{}\\
\label{eq:xachain}&=\tfrac{1}{\lambda_\alpha^{n+1}}\funof{A^{n+1}}^{\mathsf{T}}X_{\alpha}A^{n+1}+\sum_{k=0}^{n}\tfrac{\alpha}{\lambda_\alpha^{k+1}}\funof{A^{k}C}^{\mathsf{T}}CA^{k}.
\end{align}
Since for any $x_0$, $x_0^{\mathsf{T}}\funof{A^{n+1}}^{\mathsf{T}}X_{\alpha}A^{n+1}x_0\geq{}0$ and $x_0^{\mathsf{T}}\funof{A^{k}C}^{\mathsf{T}}CA^{k}x_0\geq{}0$, the right-hand side of the above implies that
\[
x_0^{\mathsf{T}}X_\alpha{}x_0\geq{}\alpha\lambda_\alpha^{-\funof{n+1}}y_n^{\mathsf{T}}y_n,
\]
where $y_n=CA^{n}x_0$ (note that $\lambda_\alpha>0$). But under the hypothesis of (ii), given any non-zero $x_0$, there exists a natural number $n$ such that $y_n\neq{}0$ (otherwise, our system would not be observable). Therefore, $x_0^{\mathsf{T}}X_\alpha{}x_0>0$, meaning that $X_\alpha{}$ is positive definite for all $\alpha>0$.

\subsection{Uniqueness of solutions}\label{sec:unique}

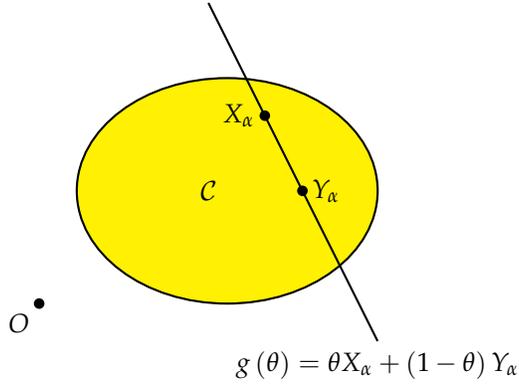
\begin{figure}
\centering
\begin{tikzpicture}
  % Draw the ellipse with a thicker boundary and fill it with yellow
  \draw[fill=yellow, thick] (1.5,1) ellipse (2cm and 1.5cm) node[left] {$\mathcal{C}$};
  
   % Define the coordinates
  \coordinate[label=below left:$O$] (O) at (-1,-.5);
  \coordinate[label=left:$X_{\alpha}$] (X) at (2,2);
  \coordinate[label=right:$Y_{\alpha}$] (Y) at (2.5,1);
  % Draw the line that extends in both directions and add the label
  \draw ($(X)!-1.5!(Y)$) -- ($(X)!3!(Y)$) node[below] {$g\funof{\theta}=\theta X_{\alpha} + \funof{1-\theta}Y_{\alpha}$};
  % Draw the points
  \fill[black] (O) circle (2pt);
  \fill[black] (X) circle (2pt);
  \fill[black] (Y) circle (2pt);
\end{tikzpicture}
\caption{Sketch of $\mathcal{C}$, two distinct points $X_\alpha,Y_\alpha\in\mathcal{C}$, and the line $\theta X_{\alpha} + \funof{1-\theta}Y_{\alpha}$. Points in the interior of $\mathcal{C}$ are positive definite, and points on the boundary are only semi-definite. Since $\mathcal{C}$ is compact, the line must pass through its boundary. This means that there is a value of $\theta$ such that $g\funof{\theta}$ is positive semi-definite, but not positive definite. It also follows that when $\theta$ becomes sufficiently large, $g\funof{\theta}\notin\mathcal{C}$. These two observations are used in \cref{sec:unique} to deduce that for each $\alpha>0$, the function $f\funof{\cdot}$ has a unique fixed point.}\label{fig:3}
\end{figure}

We will now deduce that 2) holds. We proceed by contradiction, and assume that there is at least one more fixed-point $Y_\alpha$. Let 
\[
\gamma_\alpha=\tr{\funof{A^{\mathsf{T}}Y_\alpha{}A+\alpha{}C^{\mathsf{T}}C}}
\]
and consider the function
\[
g\funof{\theta}=\theta{}X_\alpha{}+\funof{1-\theta}Y_\alpha{}.
\]
Note that although $X_\alpha$ and $Y_\alpha$ must both be positive definite by our argument from the previous subsection, $g\funof{\theta}$ is not positive definite for all $\theta$. As explained in \Cref{fig:3}, there are values of $\theta$ such that $g\funof{\theta}$ is only positive semi-definite, and when $\theta$ is large enough, $g\funof{\theta}\notin{\mathcal{C}}$. 

We will now investigate what happens when we pass matrices defined by the above function through $f\funof{\cdot}$. First, note that
\[
\begin{aligned}
A^{\mathsf{T}}g\funof{\theta}A+\alpha{}C^{\mathsf{T}}C=\theta\lambda_\alpha{}X_\alpha+\funof{1-\theta}\gamma_\alpha{}Y_\alpha\\
\implies{}f\funof{g\funof{\theta}}=\tfrac{\theta\lambda_\alpha{}}{\theta\lambda_\alpha{}+\funof{1-\theta}\gamma_\alpha{}}X_{\alpha}+
\tfrac{\funof{1-\theta}\gamma_\alpha{}}{\theta\lambda_\alpha{}+\funof{1-\theta}\gamma_\alpha{}}Y_\alpha.
\end{aligned}
\]
Observe from the right-hand-side of the above that 
\[
f\funof{g\funof{\theta}}=g\funof{\tfrac{\theta\lambda_\alpha{}}{\theta\lambda_\alpha{}+\funof{1-\theta}\gamma_\alpha{}}}.
\]
This means that when we pass a matrix of the form of $g\funof{\theta}$ through $f\funof{\cdot{}}$, we get a matrix on the same form back, and more specifically, the $\theta$ value of the returned matrix is given by
\[
g^{-1}\funof{f\funof{g\funof{\theta}}}=\tfrac{\theta\lambda_\alpha{}}{\theta\lambda_\alpha{}+\funof{1-\theta}\gamma_\alpha{}}.
\]
Therefore, when studying the fixed-points of $f\funof{\cdot}$ that can be written on the form of $g\funof{\theta}$, we can equivalently study the function $\theta\mapsto{}g^{-1}\funof{f\funof{g\funof{\theta}}}$. More specifically, $g\funof{\theta}$ is a fixed-point of $f\funof{\cdot}$ if and only if
\begin{equation}\label{eq:thetafix}
\theta=\frac{\theta\lambda_\alpha{}}{\theta\lambda_\alpha{}+\funof{1-\theta}\gamma_\alpha{}}.
\end{equation}
Suppose that $\lambda_\alpha=\gamma_\alpha$. In this case, the right-hand-side of \cref{eq:thetafix} simplifies to $\theta$, implying that $g\funof{\theta}$ is a fixed-point of $f\funof{\cdot}$ for all $\theta$. As noted earlier (and explained in \Cref{fig:3}), this implies that there exists a $\theta$ such that $g\funof{\theta}$ is positive semi-definite but not positive definite (that is, $g\funof{\theta}$ lies in $\mathcal{C}$, but not its interior). However, this is a contradiction, since in the previous subsection, we showed that all fixed points in $\mathcal{C}$ were positive definite. Therefore $\lambda_\alpha\neq{}\gamma_\alpha$. 

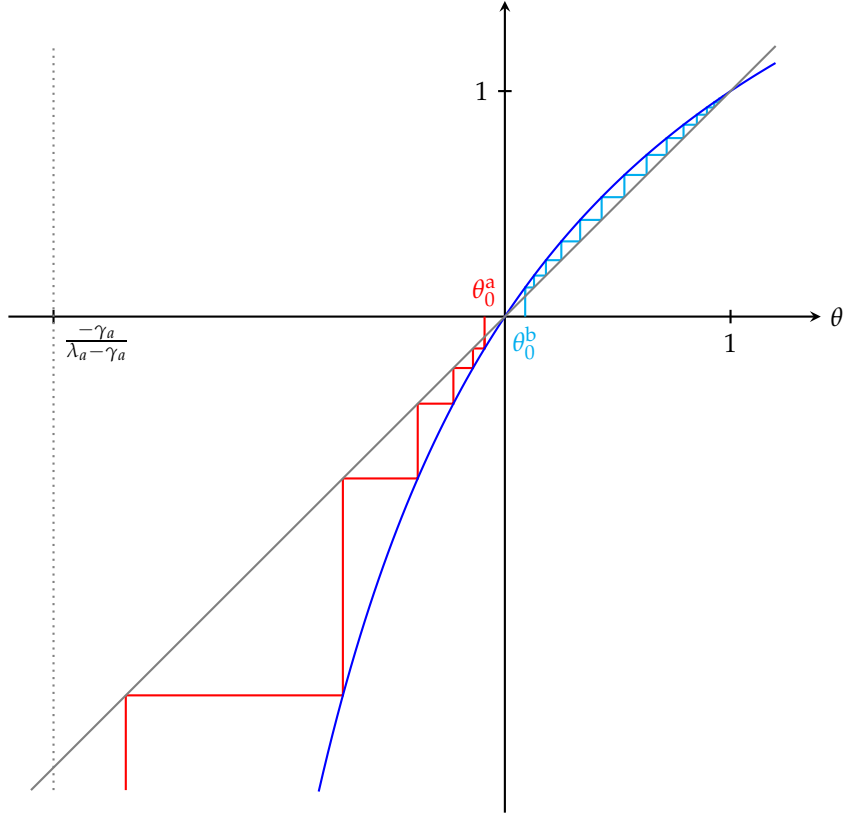
\begin{figure}
\centering
\begin{tikzpicture}[scale=3,>=stealth]  
    \pgfmathsetmacro{\aVal}{1.2}
    \pgfmathsetmacro{\bVal}{0.8}

    \draw (1,-.03) node[below] {$1$} -- (1,.03);
    \draw (-.03,1) node[left] {$1$} -- (.03,1);
    \draw ({-\bVal/(\aVal-\bVal)},-.03) node[below right,yshift=4] {$\tfrac{-\gamma_a}{\lambda_a-\gamma_a}$} -- ({-\bVal/(\aVal-\bVal)},.03);

    \draw[->](0,-2.2)--(0,1.4) node[above]{};
    \draw[->](-2.2,0)--(1.4,0) node[right]{$\theta$};
    
    \newcommand{\z}{.09}
    \draw[cyan] (\z,0) node[below] {$\theta_0^{\mathrm{b}}$} -- (\z,{\z*\aVal/(\z*\aVal+(1-\z)*\bVal)});
    
    \foreach \i in {1,...,15}{
        \pgfmathsetmacro{\w}{\z*\aVal/(\z*\aVal+(1-\z)*\bVal)}
        \draw[color=cyan](\z,\z)--(\z,\w)--(\w,\w);
        \global\let\z\w}
    
    \newcommand{\x}{-.09}
    \draw[red] (\x,0) node[above] {$\theta_0^{\mathrm{a}}$} -- (\x,{\x*\aVal/(\x*\aVal+(1-\x)*\bVal)});
    
    \foreach \i in {1,...,5}{
        \pgfmathsetmacro{\y}{\x*\aVal/(\x*\aVal+(1-\x)*\bVal)}
        \draw[color=red](\x,\x)--(\x,\y)--(\y,\y);
        \global\let\x\y}
        
    \draw[color=red] (\x,\x) -- (\x,-2.1);    
    \draw[color=blue,thick,domain=-.825:1.2,samples=100] plot (\x,{\x*\aVal/(\x*\aVal+(1-\x)*\bVal)});  
    \draw[color=gray, thick](-2.1,-2.1)--(1.2,1.2);
    \draw[color=gray, dotted]({-\bVal/(\aVal-\bVal)},-2.1)--({-\bVal/(\aVal-\bVal)},1.2);
    
\end{tikzpicture}
\caption{The blue curve shows the function $g^{-1}\funof{f\funof{g\funof{\theta}}}$, where without loss of generality, it has been assumed that $\lambda_\alpha>\gamma_\alpha$ (more specifically, $\lambda_\alpha=1.2$ and $\gamma_\alpha=0.8$). The red and cyan lines show the cobweb plots of function iterates starting at points slightly to the left and right of $\theta=0$ (the points $\theta_0^{\mathrm{a}}$ and $\theta_0^{\mathrm{b}}$ respectively). The function has a repelling fixed point at $\theta=0$ and an attracting fixed point at $\theta=1$. Since the function has an asymptote at $\theta=\tfrac{-\gamma_a}{\lambda_a-\gamma_a}$ (the dotted line), if we iteratively apply the function starting slightly to the left of the repelling fixed point, the iterates can grow arbitrarily large. This implies that for sufficiently large $n$, there exists a $\theta_0$ such that $f\funof{\theta_n}\notin\mathcal{C}$ (as $\theta_k$ varies, $f\funof{\theta_k}$ will move along the line in \Cref{fig:3} and eventually leave $\mathcal{C}$, since $\mathcal{C}$ is compact).}\label{fig:4}
\end{figure}

However if $\lambda_\alpha\neq{}\gamma_\alpha$, there are precisely two fixed points on the form $g\funof{\theta}$. As explained in \Cref{fig:4}, one must be repelling, meaning that there exists a $\theta_0$ and a natural number $n$ such that $g\funof{\theta_0}\in\mathcal{C}$ and 
\[
\theta_n=g^{-1}\funof{f^{\funof{n}}\funof{g\funof{\theta_0}}}
\]
becomes arbitrarily large. Once more, this leads to a contradiction, since this implies that given such an initial condition, for sufficiently large $n$, $g\funof{\theta_n}\not\in\mathcal{C}$ (which is absurd, since $f:\mathcal{C}\rightarrow{}\mathcal{C}$).

\subsection{Asymptotic stability implies the existence of $\lambda_\alpha=1$}

We will now deduce that 3) holds by using (i) to show that for sufficiently small $\alpha$, $\lambda_\alpha<1$. Since 
\[
\lambda_\alpha\geq{}\alpha\tr{\funof{C^{\mathsf{T}}C}}
\]
there are values of $\alpha$ such that $\lambda_\alpha>1$. As explained in \Cref{rem:4}, uniqueness of $X_\alpha$ implies that the function $\alpha\mapsto{}\lambda_\alpha$ is continuous for all $\alpha>0$, from which 3) then follows by the intermediate value theorem. We again proceed by contradiction, and instead assume that $\lambda_\alpha\geq{}1$ for all $\alpha>0$. Consider now a sequence with positive elements  $\alpha_1,\alpha_2,\alpha_3,\ldots{}$ such that $\alpha_n\rightarrow{}0$, and let $X_{\alpha_{n_k}}$ be any convergent subsequence such that $X_{\alpha_{n_k}}\rightarrow{}X^*$ (which must exist by the Bolzano-Weierstrass theorem). It follows that $X^*$ is non-zero and positive semi-definite, 
\[
\lambda^*=\tr\funof{A^{\mathsf{T}}X^*A}\geq{}1\text{, and}\,\lambda^*X^*=A^{\mathsf{T}}X^*A.
\]
This implies that for all $x$, $x^{\mathsf{T}}A^{\mathsf{T}}X^*Ax=\lambda^*x^{\mathsf{T}}X^*x\geq{}x^{\mathsf{T}}X^*x$. Therefore, along any trajectory $x_n=A^nx_0$, 
\[
x_n^{\mathsf{T}}X^*x_n\geq{}x_0^{\mathsf{T}}X^*x_0.
\]
This would imply that for any initial condition such that $x_0^{\mathsf{T}}X^*x_0>0$ (which must exist, since $X^*$ is non-zero and positive semi-definite), $x_n$ does not tend to zero, contradicting the hypothesis (i) of asymptotic stability.

\begin{remark}\label{rem:2}
We have not proved the implications
\[
\text{(ii)}\land\text{(iii)}\implies{}\text{(i)}\quad\text{and}\quad\text{(i)}\land\text{(iii)}\implies{}\text{(ii)}.
\]
Why not give these a go yourself? Here are some hints for tackling these steps in a way that avoids linear algebra tricks.
\begin{enumerate}
\item Making use of the Lyapunov candidate 
\[
V\funof{x}=x^{\mathsf{T}}Qx
\]
where $Q$ is the positive definite solution to the Lyapunov equation.
\item For the implication $\text{(ii)}\land\text{(iii)}\implies{}\text{(i)}$, use LaSalle's invariant set principle.
\item For the implication $\text{(i)}\land\text{(iii)}\implies{}\text{(ii)}$, note that asymptotic stability implies that there exists a natural number $n$ such that $V\funof{x_0}-V\funof{x_n}>0$. Try using \cref{eq:xachain} to deduce observability.
\end{enumerate}
\end{remark}

\begin{remark}\label{rem:4}
We will now explain why having a unique solution to the fixed-point equation
\[
X_\alpha=\frac{A^{\mathsf{T}}X_\alpha{}A+\alpha{I}}{\tr\funof{A^{\mathsf{T}}X_\alpha{}A+\alpha{I}}}
\]
for each $\alpha>0$ ensures that the function $\alpha\mapsto{}\lambda_\alpha$ is continuous for positive $\alpha$. 
First, note that uniqueness of $X_\alpha$ ensures that the function $\alpha\mapsto{}X_\alpha$ is well defined. Continuity of this function then follows from continuity of $h:\funof{0,\infty}\times{}\mathcal{C}\rightarrow{}\mathcal{C}$, where
\[
h\funof{\alpha,X}=X-\frac{A^{\mathsf{T}}XA+\alpha{}C^{\mathsf{T}}C}{\tr{\funof{A^{\mathsf{T}}XA+\alpha{}C^{\mathsf{T}}C}}}
\]
and $\mathcal{C}$ is the set of positive semi-definite matrices with unit trace [see \cref{eq:unittr}]. To see this, let $\alpha_n\rightarrow{}\alpha^*$, where $\alpha^*\in\funof{0,\infty{}}$, and $X_{\alpha_{n_k}}$ be any convergent subsequence such that $X_{\alpha_{n_k}}\rightarrow{}X^*$ (which must exist by the Bolzano-Weierstrass theorem). Since 
\[
h\funof{\alpha_{n_k},X_{\alpha_{n_k}}}=0\;\;\text{and}\;\;\funof{\alpha_{n_k},X_{\alpha_{n_k}}}\rightarrow{}\funof{\alpha^*,X^*}
\]
continuity of $h\funof{\cdot,\cdot}$ ensures that $h\funof{\alpha^*,X^*}=0$. Uniqueness then ensures that $X^*=X_{\alpha^*}$, and so for any $\alpha^*\in\funof{0,\infty}$, $\alpha_n\rightarrow{}\alpha^*$ implies that $X_{\alpha_n}\rightarrow{}X_{\alpha^*}$ [or in other words, the function $\alpha\mapsto{}X_\alpha$ is continuous on $\funof{0,\infty}$]. Continuity of $\alpha\mapsto{}\lambda_\alpha$ then follows immediately, since $\lambda_\alpha$ depends continuously on $\alpha$ and $X_\alpha$.
\end{remark}

\bibliographystyle{IEEEtran}
\bibliography{references.bib}

\begin{thebibliography}{1}
\providecommand{\url}[1]{#1}
\csname url@rmstyle\endcsname
\providecommand{\newblock}{\relax}
\providecommand{\bibinfo}[2]{#2}
\providecommand\BIBentrySTDinterwordspacing{\spaceskip=0pt\relax}
\providecommand\BIBentryALTinterwordstretchfactor{4}
\providecommand\BIBentryALTinterwordspacing{\spaceskip=\fontdimen2\font plus
\BIBentryALTinterwordstretchfactor\fontdimen3\font minus
  \fontdimen4\font\relax}
\providecommand\BIBforeignlanguage[2]{{%
\expandafter\ifx\csname l@#1\endcsname\relax
\typeout{** WARNING: IEEEtran.bst: No hyphenation pattern has been}%
\typeout{** loaded for the language `#1'. Using the pattern for}%
\typeout{** the default language instead.}%
\else
\language=\csname l@#1\endcsname
\fi
#2}}

\bibitem{PW98}
J.~W. Polderman and J.~C. Willems, \emph{Introduction to Mathematical Systems
  Theory: A Behavioral Approach}.\hskip 1em plus 0.5em minus 0.4em\relax
  Springer Science \& Business Media, 1998, no.~26.

\bibitem{AZ18}
M.~Aigner and G.~M. Ziegler, \emph{Pigeon-hole and double counting}.\hskip 1em
  plus 0.5em minus 0.4em\relax Berlin, Heidelberg: Springer Berlin Heidelberg,
  2018, pp. 195--205.

\bibitem{Gal79}
D.~Gale, ``The game of {H}ex and the {B}rouwer fixed-point theorem,'' \emph{The
  American Mathematical Monthly}, vol.~86, no.~10, pp. 818--827, 1979.

\bibitem{DM21}
G.~Dinca and J.~Mawhin, \emph{History of the Brouwer Fixed Point
  Theorem}.\hskip 1em plus 0.5em minus 0.4em\relax Cham: Springer International
  Publishing, 2021, pp. 391--412.

\bibitem{Mac00}
C.~R. MacCluer, ``The many proofs and applications of {P}erron's theorem,''
  \emph{SIAM Review}, vol.~42, no.~3, pp. 487--498, 2000.

\bibitem{Par00}
P.~Parrilo, ``Structured semidefinite programs and semialgebraic geometry
  methods in robustness and optimization,'' Ph.D. dissertation, California
  Institute of Technology, 2000.

\bibitem{RV18}
A.~Rantzer and M.~E. Valcher, ``A tutorial on positive systems and large scale
  control,'' in \emph{2018 IEEE Conference on Decision and Control (CDC)},
  2018, pp. 3686--3697.

\end{thebibliography}

\end{document}